\documentstyle[12pt]{article}
\begin{document}
\pagestyle{plain}
\title{\bf To the nonlinear quantum mechanics}
\author{\bf Miroslav Pardy \\
Department of Physical Electronics \\
Faculty of Science, Masaryk University \\
Kotl\'{a}\v{r}sk\'{a} 2, 61137 Brno, Czech Republic\\
Email:pamir@physics.muni.cs}
\date{\today}
\maketitle

\vspace{50mm}

\begin{abstract}
The Schr\"{o}dinger equation with the nonlinear term $-b(\ln|\Psi|^2)  \Psi $ is
derived by the natural generalization of the hydrodynamical model of quantum
mechanics. The nonlinear term appears to be logically necessary because it
enables explanation of the classical limit of the wave function, the collaps of the
wave function and solves the Schr\"{o}dinger cat paradox.
\end{abstract}

\newpage
\section{Introduction}
\hspace{3ex}
Many authors have suggested that the quantum mechanics based on linear
Schr\"{o}dinger equation is only an approximation of some more nonlinear theory
with the nonlinear Schr\"{o}dinger equation. The motivation for considerring
the nonlinear quations is to get some more nonstandard solution in order to get
the better understanding of the synergism of wave and particle.

The ambicious program to create nonlinear wave mechanics was elaborated by
de~Broglie [1] and his group.
Bialynicki-Birula and Mycielski [2]
considered the generalized Schr\"{o}dinger equation with the additional term
$F(|\Psi|^2)\Psi$  where $F$ is some arbitrary function which they later
specified to $-b(\ln|\Psi|^2)$, $b > 0$. The nonlinear term was selected
by assuming the factorization of the wave function for the composed system.

The most attractive feature of the logarithmic nonlinearity is the existence of
the lower energy bound and validity of Planck's relation $E=\hbar\omega$. At
the same time the Born interpretation of the wave function cannot be changed.
In this theory the estimation of $b$ was given by the relation
$b<4\times10^{-10}eV$ following from the agreement between theory and
the observed $2S-2P$ Lamb shift in hydrogen. This implies an upper bound
to the electron soliton spatial width of 10 $\mu$m.

Shimony [3] proposed an experiment which is based on idea that a phase shift
occurs when an absorber is moved from one point to another along the path of
one of the coherent split beams in a neutron interferometer. In case of the
logarithmic nonlinearity Shull at al. [4] performed the experiment with
a~ two-crystal interferometer. They searched for a phase shift when
an attenuator was moved along the neutron propagation direction in
one arm of the interferometer. A sheet of Cd, 0.086 mm thick, was used
for the absorber.They obtained the upper bound on $b$
of~$3.4\times10^{-13}eV$ which is more than three orders of magnitude smaller
than the bound estimated by Bialynicky-Birula and Mycielski [2].

The best upper limit on $b$ has been reported by G\"{a}hler, Klein and
Zeilinger [5] who has been  searched for variations
in the free space propagation of neutrons.
20~{\AA} neutrons were difracted from an abrupt highly absorbing
knife edge at the object position. By comparing the experimental
results with the solution to the ordinary Schr\"{o}dinger equation they
were able to get the limit  $b<3\times 10^{-15}eV$, which corresponds
to an alectron soliton width of 3 mm. The similar results was obtained
by the same group from diffraction a 100 $\mu$m boron wire.

To our knowledge the M\"{o}ssbauer effect was not used to determine the
constant $b$ although this effect allows to measure energy losses
smaller than $10^{-15}eV$. Similarly the Josephson effect has been not applied
for the determination of the constant $b$.

We see that the constant $b$ is very small, nevertheless we cannot it neglect
a~priori, because we do not know its role in the future physics. The
corresponding analogon is the Planck constant which is also very
small, however, it plays the fundamental role in physics.

The goal of this article is to give the new derivation of the logarithmic
nonlinearity, to find the solution of the nonlinear Schr\"{o}dinger equation
of the one-dimensional case and to show that in the mass
limit $ m\rightarrow\infty$
we get exactly the delta-function behavior of the probability of finding the
particle at point $x$. It means that there exists the classical motion of a
particle with sufficient big mass. The nonlinearity of the Schr\"{o}dinger
equation also solves the collaps of the wave function and the Schr\"{o}dinger
cat paradox. We will start from the hydrodynamical formulation of quantum
mechanics. The mathematical generalization of the Euler hydrodynamical
equations leads automatically to the logarithmic term with $b>0$.

\section{The derivation of the nonlinear Schr\"{o}dinger equation}
\hspace{3ex}
We respect here the so called Dirac heuristical principle [6]
according to which it is useful to postulate some mathematical requirement in
order to get the true information about nature. While the mathematical
assumption is intuitive, the consequences have the physical interpretation or
in other words they are physically meaningfull. In derivation of the
logarithmic nonlinearity we use just the Dirac method.

According to Madelung [7], Bohm and Vigier [8],
Wilhelm [9], Rosen [10] and
others, the original Schr\"{o}dinger equation can be transformed into the
hydrodynamical system of equations by using the so called Madelung ansatz:
\begin{equation}
\label{1}
\Psi={\sqrt n}\*e^{\frac{i}{\hbar}\*S},
\end{equation}
where $n$ is interpreted as the density of particles and $S$ is the classical
action for $\hbar\rightarrow 0$. The mass density is defined by relation
$\varrho=n\*m$ where $m$ is mass of~a~particle.

It is well known that after insertion of the relation (\ref{1}) into the
original Schr\"{o}dinger equation
\begin{equation}
\label{2}
i \hbar \frac {\partial \Psi}{\partial t} =
 -  \frac {\hbar^2}{2m}\*\Delta \Psi + V\* \Psi,
\end{equation}
where $V$ is the potential energy, we get, after separating the real and
imaginary parts, the following system of equations:
\begin{equation}
\label{3}
\frac {\partial S}{\partial t} - \frac {1}{2m}\* (\nabla\* S)^2 + V =
\frac {\hbar^2}{2m} \frac {\Delta \sqrt{n}}{\sqrt{n}}
\end{equation}
\begin{equation}
\label{4}
\frac {\partial n}{\partial t} + {\rm div}(n\*{\bf v}) = 0
\end{equation}
with
\begin{equation}
\label{5}
{\bf v}=\frac {\nabla S}{m}.
\end{equation}

Equation (\ref{3}) is the Hamilton-Jacobi equation with the additional term
\begin{equation}
\label{6}
V_q = - \frac {\hbar^2}{2m} \frac {\Delta \sqrt{n}}{\sqrt{n}},
\end{equation}
which is called the quantum Bohm potential and equation (\ref{4}) is the
continuity equation.

After application of operator $\bigtriangledown$ on eq. (\ref{3}), it can
be cast into the Euler hydrodynamical equation of the form:
\begin{equation}
\label{7}
\frac{\partial {\bf v}}{\partial t}+({\bf v} \cdot \nabla)\* {\bf v}=
- \frac {1}{m}\*\nabla\* (V+V_q).
\end{equation}
It is evident that this equation is from the hydrodynamical point of view
incomplete as a consequence of the missing term
$-\varrho^{-1}\*\nabla\* p$
where $p$ is hydrodynamical pressure. We use here this fact just as the crucial
point for derivation of the nonlinear Schr\"{o}dinger equation. We complete the
equation (\ref{7}) by adding the pressure term and in such a way we get the
total Euler equation in the form:
\begin{equation}
\label{8}
m \left( \frac{\partial {\bf v}}{\partial t} + ({\bf v}\cdot\nabla)\*
{\bf v}\right)= - \nabla\*(V+V_q)- \nabla\*F,
\end{equation}
where
\begin{equation}
\label{9}
\nabla\*F = \frac{1}{n}\* \nabla p.
\end{equation}

The equation (8) can be obtained by the Madelung procedure from the
following extended Schr\"{o}dinger equation
\begin{equation}
\label{10}
i \hbar \frac {\partial \Psi}{\partial t} =
 -  \frac {\hbar^2}{2m}\*\Delta \Psi + V\* \Psi + F \Psi
\end{equation}
on the assumption that it is possible to determine $F$ in term of the wave
function. From the vector analysis follows that the necessary condition of the
existence of $F$ as the solution of the equation (\ref{9}) is $ rot\;
 grad\; f = 0$, or,
\begin{equation}
\label{11}
{\rm rot}(n^{-1}\* \nabla \* p) =0,
\end{equation}
which enables to take the linear solution in the form
\begin{equation}
\label{12}
p= -bn = -b|\Psi|^2,
\end{equation}
where $b$ is some arbitrary constant.  We do not consider
the more general solution of eq. (11). Then, from eq. (\ref{9}) i.e.
$ {\rm grad}\;F={\bf a}$ we have:
\begin{equation}
\label{13}
F= \int{a_i}\,dx_i = -b\* \int{\frac{1}{n}\,dn} = -b\* \ln|\Psi|^2,
\end{equation}
where we have omitted the additive constant which plays no substantial role
in the Schr\"{o}dinger equation.

Now, we can write the generalized Schr\"{o}dinger equation which correspond
to the complete Euler equation (\ref{8}) in the following form:
\begin{equation}
\label{14}
i \hbar \frac{\partial \Psi}{\partial t}= - \frac{\hbar^2}{2m}\* \Delta
\* \Psi + V\Psi - b\*(\ln{|\Psi|^2})\* \Psi .
\end{equation}

Let us approach the solving the equation (\ref{14}).

\section{The soliton-wave solution of the nonlinear
Schr\"{o}dinger equation}
\hspace{3ex}
Let be $c, ({\rm Im} \;c =0), v, k, \omega$ some parameters and let us insert function
\begin{equation}
\label{15}
\Psi (x,t)= c\* G (x-v\*t)\* e^{i\*k\*x-i\*\omega\*t}
\end{equation}
into the one-dimensional equation (\ref{14}) with $V=0$. Putting the imaginary
part of the new equation to zero, we get
\begin{equation}
\label{16}
v= \frac{\hbar\*k}{m}
\end{equation}
and for function $G$ we get the folloving nonlinear equation (symbol ' denotes
derivation with respect to $\xi= x-vt)$:
\begin{equation}
\label{17}
G'' + A\*G + B(\ln{G})G = 0,
\end{equation}
where
\begin{equation}
\label{18}
A= \frac{2m}{\hbar}\*\omega - k^2 + \frac{2m}{\hbar^2}\*b \*\ln{c^2}
\end{equation}
\begin{equation}
\label{19}
B= \frac{4mb}{\hbar^2}.
\end{equation}

After multiplication of eq. (\ref{17}) by $G'$ we get:
\begin{equation}
\label{20}
\frac{1}{2}\*{\left[ G'^2 \right]}^{'} + \frac{A}{2}\*{\left[ G^2\right]}^{'}
+  B\* {\left[ \frac{G^2}{2} \ln{G} - \frac{G^2}{4} \right]}^{'} =0,
\end{equation}
or, after integration
\begin{equation}
\label{21}
G'^2=- AG^2 - BG^2 \ln{G} + \frac{B}{2}\* G^2 + const.
\end{equation}

If we choose the solution in such a way that $G(\infty)=0$ and
$G'(\infty)=0$, we get $const.=0$ and after elementary operations we get the
following differential equation to be solved:
\begin{equation}
\label{22}
\frac{dG}{G \sqrt{a-B\*\ln{G}}}= d\xi,
\end{equation}
where
\begin{equation}
\label{23}
a=\frac{B}{2} - A.
\end{equation}

Equation (\ref{22}) can be solved by the elementary integration and the
result is
\begin{equation}
\label{24}
G= e^{\frac{a}{B}}\*e^{-\frac{B}{4}\*(\xi+d)^2},
\end{equation}
where $d$ is some constant.

The corresponding soliton-wave function is evidently in the one-dimensional
free particle case of the form:
\begin{equation}
\label{25}
\Psi(x,t)=
 c\*e^{\frac{a}{B}}\*e^{-\frac{B}{4}\*(x-vt+d)^2}\*e^{ikx-i\omega\*t}.
\end{equation}

\section{Normalization and the classical limit}
\hspace{3ex}
It is not necessary to change the standard probability interpretation of the
wave function. It means that the normalization condition in our one-dimensional
case is
\begin{equation}
\label{26}
\int_{-\infty}^{\infty}{\Psi^*\*\Psi\,dx} =1.
\end{equation}
Using the Gauss integral
\begin{equation}
\label{27}
\int_{0}^{\infty}{e^{-\lambda^2\*x^2}\,dx} =
\frac{\sqrt{\pi}}{2\lambda},
\end{equation}
we get with $\lambda= {\left(\frac{B}{2}\right)}^{\frac{1}{2}}$
\begin{equation}
\label{28}
c^2\*e^{\frac{2a}{B}}= {\left(\frac{B}{2\pi}\right)}^{\frac{1}{2}}
\end{equation}
and the density probability $\Psi^*\Psi = \delta_m(\xi) $ is of the form
(with $d=0$):
\begin{equation}
\label{29}
\delta_m(\xi)= \sqrt{\frac{m\alpha}{\pi}}\* e^{-\alpha m \xi^2}
\hspace{5mm};
\hspace{5mm}
\alpha= \frac{2b}{\hbar^2}.
\end{equation}

It may be easy to see that $\delta_m(\xi)$ is the delta-generating function and
for $m \rightarrow \infty$ is just the Dirac $\delta$-function.

It means that the motion of a particle with sufficiently big mass $m$ is
strongly localized and in other words it means that the motion of this particle
is the classical one. Such behaviour of a particle cannot be obtained in the
standard quantum mechanics because the plane wave
\begin{equation}
\label{30}
e^{ikx-i\omega\*t}
\end{equation}
corresponds to the free particle with no possibility of localization
for $m \rightarrow \infty$.

Let us still remark that coefficient $c^2$ is real and positive number because
it is a result of the solution of equation (\ref{28}) which can be
transformed into equation $(x=c^2)$
\begin{equation}
\label{31}
x^{1-r}= constant.
\end{equation}

\section{The collaps of the wave function}
\hspace{3ex}
The nonlinear quantum mechanics explains by the easy way the collaps or the
reduction of the wave function. Let us suppose that an electron
is impinging on the screen. It means in other words that it is captured by
a system of atoms in the screen. However, the system of a great amount of
particles has mass $M$ which is enormous in comparison with the mass of
electron. It means that the electron together with the system of particles have
mass $M+m$ which is practically sufficiently big in order to get the
soliton-wave solution of the form which describes the strong
localization of the system consisting of electron and surrounding particles.
The classical analogon of this situation is obviously the ballistic
pendulum. Such localization is not possible in the standard quantum mechanics
of the known textbooks. Therefore the nonlinear quantum mechanics of the above
type with the logarithmic nonlinearity solves one of the old problems of the
standard quantum mechanics.

\section{The principle of superposition and the Schr\"{o}\-din\-ger~cat}
\hspace{3ex}
If $\varphi_1$ and $\varphi_2$ are two different solution of the
nonlinear Schr\"{o}dinger  equation then the linear combination
$\varphi=a\*\varphi_1 + b\*\varphi_2$ where $a$ and $b$ are the arbitrary
constants is not the solution of the same equation because of its nonlinearity.
In other words the original principle of superposition of the
standard quantum mechanics is broken. The consequence of the breaking of the
principle of superposition is the resolution of the
Schr\"{o}dinger cat paradox.

In the cat paradox [11], Schr\"{o}dinger
imagined the arrangement in which
a cat is confined to a box which containes a lethal device that may be either
triggered or left passive according to whether radioactive nucleus decays or
fails do decay. If the radioactive decay takes place, which he assumed would
happen half the time, a hammer would strike a vial of cyanide and cat would
be dispatched. It means that the quantum signal is amplified in order
to be possible to move by some equipment the hammer.

In the mathematical form, to the microstate $\varphi_a$ corresponds the
macrostate $\varphi_A$ where the microstate is the decay of the nucleus and
the macrostate is the striking hammer  a vial cyanide. If to the nondecay
of the nucleus corresponds the state $\varphi_b$ then according to
Schr\"{o}dinger the corresponding macrostate is $\varphi_B$. In the standard
quantum mechanics it is possible the superposition $\varphi_a+\varphi_b$.
To this state corresponds the superposition $\varphi_A+\varphi_B$, which we can
interprete as the cat is partly dead and partly alive, which is impossible
because the vial of cyanide cannot be partly broken and partly unbroken.

The resolution of this contradiction in the nonlinear quantum mechanics
with the logarithmic nonlinearity is that the superposition of the
states is not solution of the nonlinear Schr\"{o}dinger equation and therefore
the vial is or broken or not. The answer is unambiguous.

\section{Discussion}
\hspace{3ex}
We have seen that the introduction of the logarithmic nonlinearity in the
Schr\"{o}dinger equation was logically supported by the fact that the nonlinear
Schr\"{o}dinger equation gives results which are physically meaningful.
We have obtained the correct mass limit of the wave function and explained
the collaps of it. The Schr\"{o}dinger cat paradox was also explained.

The further strong point of the nonlinear Schr\"{o}dinger equation (14)
is the result (16) which is equivalent to the famous de Broglie relation

\begin{equation}
\label{32}
\lambda = \frac{h}{p}
\end{equation}
because of $\lambda = 2\pi/k = 2\pi(\hbar/mv) = 2\pi(h/2\pi)(1/p)$
and it means that de Broglie relation is involved in this form of the
nonlinear quantum mechanics.

The nonlinear equation (14) has also the normalized plane-wave solution
\begin{equation}
\label{33}
\Psi(x,t) = \frac{1}{\sqrt2\pi}\*e^{ikx-i\omega t}.
\end{equation}
After insertion of eq. (33) into eq. (14), we get the following dispersion
relation:

\begin{equation}
\label{34}
\hbar\omega = \frac {\hbar^{2}k^{2}}{2m} + b\ln(2\pi),
\end{equation}
from which the relations follows:

\begin{equation}
\label{35}
\hbar\omega = b\ln(2\pi);\quad k = 0
\end{equation}
and

\begin{equation}
\label{36}
k = \pm i \sqrt{\frac {2m}{\hbar^{2}}b\ln(2\pi)}; \quad \omega = 0.
\end{equation}

It is no easy to give the physical interpretation of eqs. (35) and (36)
and so we cannot say that the plane-solution of the nonlinear Schr\"{o}dinger
equation is physically meaningful. Only the soliton-wave solution
of the nonlinear Schr\"{o}dinger equation can be taken as relevant.
Only this solution
is suitable for the physical verification.
The possible new tests of the nonlinear quantum mechanics are discussed in
the author article [12].

The generalization to the motion of particle in the electromagnetic field
with potentials $\varphi({\bf x},t)$ and ${\bf A}({\bf x},t)$ can be performed
by the standard transformation
\begin{equation}
\label{37}
\frac{\hbar}{i}\*\nabla \rightarrow
\frac{\hbar}{i}\*\nabla -
\left(\frac{e}{c}\right)\* {\bf A}\*({\bf x},t)
\end{equation}
and adding the scalar potential energy $e\varphi(x,t)$ in the
Schr\"{o}dinger equation for the free particles. According to [2] the
solution of the equation in this case can be taken in the form
\begin{equation}
\label{38}
\Psi({\bf x},t)= e^{\frac{i}{\hbar}\*S} \* G({\bf x}-{\bf u}(t)),
\end{equation}
where function $G$ is necessary to determine. In the similar form
the problem was yet solved [13].

Kamesberger and Zeilinger [14] have given the numerical
solution of the original Schr\"{o}dinger equation and this equation
with the nonlinear term  $-b(\ln|\Psi|^2) \Psi $  in order to
visualize the spreading of the diffractive waves. When comparing the
evolution patterns of the nonlinear case with the linear one, one
notices that the maxima are more pronounced in the nonlinear solution.
It can be understood as a mechanism compressing the wave maxima
spatially. In the quantitative comparison of the both cases this
enhacement of the maxima and minima can be seen very clearly.

Although we have given reasons for the introducing of the nonlinear
Schr\"{o}dinger equation it is obvious that only the crucial experiments
can establish the physical and not only logical necessity of such equation.
In case that the nonlinear Schr\"{o}dinger equation will be confirmed by
experiment, then it can be expected that it will influence other parts of
theoretical physics.

\end{document}